\documentclass[a4paper,fleqn]{cas-sc}

\usepackage[authoryear,longnamesfirst]{natbib}

\def\tsc#1{\csdef{#1}{\textsc{\lowercase{#1}}\xspace}}
\tsc{WGM}
\tsc{QE}

\begin{document}
\let\WriteBookmarks\relax
\def\floatpagepagefraction{1}
\def\textpagefraction{.001}

\shorttitle{Particle Acceleration in the Pre-Structure Universe}    

\shortauthors{Ha} 

\title [mode = title]{Stochastic Particle Acceleration during Pressure-Anisotropy-Driven Magnetogenesis in the Pre-Structure Universe}  



%

\author[1]{Ji-Hoon Ha}[orcid=0000-0001-7670-4897]

\cormark[1]

\fnmark[1]

\ead{hjhspace223@gmail.com}


\credit{Conceptualization, Methodology, Formal analysis, Investigation, Writing – original draft, Writing – review \& editing.}

\affiliation[1]{organization={Korea Astronomy and Space Science Institute},
            addressline={776 Daedeok-daero, Yuseong-gu}, 
            city={Daejeon},
            postcode={34055}, 
            country={Republic of Korea}}

\cortext[1]{Corresponding author}


\begin{abstract}
We investigate whether stochastic acceleration associated with pressure-anisotropy-driven magnetogenesis can generate a dynamically significant population of cosmic rays (CRs) prior to nonlinear structure formation.
As magnetic fields amplify in the early Universe, the associated increase in gyrofrequency enhances pitch-angle scattering, potentially shortening the stochastic acceleration time.
We derive an analytic criterion for efficient cosmological acceleration by comparing the acceleration timescale with the Hubble time, which defines a critical magnetic field and a corresponding CR turn-on redshift $z_{\rm on}$.
For representative parameters, we find $z_{\rm on}\sim1.7$.
To quantify the resulting particle population, we solve a Fokker-Planck equation for the isotropic ion (proton) distribution in the redshift interval $z=10\rightarrow z_{\rm on}$, including Coulomb energy losses in a fully ionized intergalactic medium.
Throughout most of this epoch, adiabatic expansion dominates over stochastic energization, and Coulomb cooling efficiently thermalizes low-energy particles, introducing an effective low-energy threshold at energies of order ${\mathcal O}(10)$ keV.
As a result, the distribution remains close to a cooling Maxwellian, and the formation of a suprathermal tail is strongly suppressed even in the presence of a pre-existing nonthermal component.
Even under optimistic assumptions corresponding to the strong-scattering limit, the maximum attainable ion energy reaches at most $\mathcal{O}(10^2)$ GeV.
These results indicate that efficient CR production in the intergalactic medium is intrinsically tied to the onset of structure-formation shocks, while earlier microinstability-driven stochastic processes can provide at most a modest pre-acceleration.
\end{abstract}




\maketitle

\section{Introduction}
\label{sec:s1}

Cosmic rays (CRs) are ubiquitous nonthermal components of the Universe, observed across environments ranging from galaxies to galaxy clusters. 
In the Milky Way, the CR energy density is of order $\sim 1~{\rm eV\,cm^{-3}}$, comparable to that of the thermal interstellar medium, implying that CRs play a dynamically significant role in galactic systems \citep[e.g.,][]{Murase2019}.
It is widely accepted that supernova remnants accelerate CRs up to $\sim 10^{15}$ eV via diffusive shock acceleration (DSA) \citep[e.g.,][]{Bell1978,Blandford1978,Drury1983}, and multiwavelength observations provide strong evidence for relativistic particle acceleration at such shocks \citep[e.g.,][]{Koyama1995,Ohira2011,Ackermann2013}.
In some cases, CR precursors predicted by DSA have been observationally identified \citep{Ohira2017}.
In the context of large-scale structure formation, DSA at collisionless shocks is likewise regarded as a primary mechanism responsible for accelerating particles to relativistic energies.
Structure-formation shocks with velocities $u_s \sim 10^2$--$10^3~{\rm km\,s^{-1}}$ naturally arise during hierarchical clustering, and numerical simulations support their role in generating CR populations in the intracluster and intergalactic media \citep[e.g.,][]{Miniati2000,Ryu2003,Pfrommer2006,Hoeft2008,Skillman2008,Vazza2009,Hong2014,Schaal2015,Ha2018, Ha2020, Ha2023}.

However, an important open question concerns the origin of the first nonthermal particles.
Cosmological simulations indicate that a large fraction of gas thermalization and cosmic-ray production occurs at structure-formation shocks emerging at relatively low redshifts ($z \lesssim {\rm a~few}$) \citep[e.g.,][]{Ryu2003}.
If CR production is tied exclusively to such shocks, then a pre-existing suprathermal seed population may not be required. 
Nevertheless, energetic events at earlier epochs may still generate the first CRs.
For instance, early shock-driven scenarios associated with the first supernova explosions at $z \sim 20$ have been proposed as possible sources \citep[e.g.,][]{Miniati2011,Ohira2019}. 
Another possibility arises during the subsequent pre-structure epoch, before large-scale structure formation enters the nonlinear regime and cosmological shocks become widespread ($z \lesssim \text{a few}$).
During this period, the Universe undergoes a prolonged phase of weak magnetization, and magnetic fields are expected to grow gradually through various amplification processes \citep[e.g.,][]{Ryu2008,Widrow2012}.
If weak but cumulative acceleration processes operate during this intermediate redshift interval ($z \sim 10 \rightarrow {\rm a~few}$), they could establish a low-level CR background that modifies subsequent injection and acceleration at structure-formation shocks.
Determining whether such a pre-shock population exists is therefore relevant for understanding both the efficiency of DSA and the thermal history of the early intergalactic medium.

In parallel, recent studies of cosmological magnetogenesis have emphasized the role of pressure-anisotropy-driven plasma instabilities in amplifying magnetic fields in weakly magnetized environments \citep[e.g.,][]{Schekochihin2005,Schekochihin2006a,Schekochihin2006b, Falceta2015, Ha2025}.
As the magnetic field grows, the associated ion (proton) gyrofrequency increases, enhancing pitch-angle scattering and potentially reducing the characteristic acceleration time of stochastic (second-order Fermi) processes operating in turbulence \citep[e.g.,][]{Fermi1949, Brunetti2007, Petrosian2012}.
This raises a natural question: {\it can instability-assisted stochastic acceleration during cosmological magnetogenesis generate a dynamically significant CR population prior to the emergence of structure-formation shocks?}

In this work, we investigate this possibility by combining (i) an analytic acceleration-time criterion based on the comparison between the stochastic acceleration time and the Hubble time, and (ii) a Fokker-Planck model for the isotropic ion distribution.
This approach allows us to quantify both the redshift at which stochastic acceleration becomes cosmologically viable and the magnitude of any resulting nonthermal population.
The central question we address is whether instability-enhanced scattering during cosmological magnetogenesis can reduce the stochastic acceleration time below the cosmological expansion timescale over any redshift interval prior to the widespread emergence of structure-formation shocks. 
In particular, we derive a critical condition for cosmologically viable second-order Fermi acceleration and follow the redshift evolution of the ion distribution under the combined effects of stochastic diffusion and adiabatic expansion.
This framework enables a self-consistent assessment of (i) the epoch at which stochastic acceleration may become competitive with cosmic expansion, and (ii) the magnitude of any resulting suprathermal population.
Energy losses due to Coulomb interactions are known to play an important role in shaping the low-energy spectrum of charged particles in diffuse astrophysical plasmas \citep[e.g.,][]{Gould1971,Sarazin1999, Petrosian2012}. In particular, Coulomb collisions can efficiently thermalize sub-relativistic particles and suppress the formation of nonthermal tails in weakly collisional environments. However, their impact on stochastic acceleration in the cosmological context—especially during the magnetogenesis phase prior to nonlinear structure formation—has not been systematically quantified. In this work, we incorporate Coulomb losses into both the analytic framework and the Fokker--Planck model, and examine how collisional cooling modifies particle injection and the resulting nonthermal population.

We emphasize that the present work does not model fast injection mechanisms, such as reconnection-driven acceleration in turbulent current sheets
\citep[e.g.,][]{Comisso2018,Comisso2019}. 
Such processes may inject particles from the thermal pool into the suprathermal regime on timescales shorter than the stochastic acceleration timescale considered here. 
Our results therefore constrain the efficiency of the stochastic acceleration channel associated with instability-mediated scattering, rather than excluding all possible sources of suprathermal seed particles prior to structure formation.

The paper is organized as follows.
In Section~\ref{sec:s2}, we derive the stochastic acceleration timescale within the instability-mediated magnetogenesis framework and obtain an analytic criterion for the cosmological viability of stochastic acceleration by considering adiabatic losses associated with cosmic expansion.
This provides an optimistic upper bound on the efficiency of particle acceleration and defines a critical magnetic field and a corresponding turn-on redshift.
In Section~\ref{sec:s3}, we extend this analysis by including Coulomb energy losses in a fully ionized intergalactic medium. 
In Section~\ref{sec:s4}, we solve a Fokker--Planck equation to quantify the resulting ion distribution.
This allows us to assess the impact of collisional cooling on particle injection and to identify an effective low-energy threshold for stochastic acceleration.
In Section~\ref{sec:s5}, we compare the stochastic acceleration timescale with that of DSA in the Bohm-diffusion limit.
Finally, Section~\ref{sec:s6} summarizes our results and discusses their cosmological implications.

\section{Cosmological Viability of Stochastic Acceleration During Pressure-Anisotropy-Driven Magnetogenesis}
\label{sec:s2}

Magnetic-field amplification mediated by pressure-anisotropy-driven instabilities can substantially modify particle transport properties in weakly magnetized plasmas. Since magnetic fields play a central role in regulating particle scattering, the growth of cosmic magnetic fields prior to the onset of large-scale structure formation may provide conditions for stochastic particle acceleration.
In this section, we assess the cosmological viability of such acceleration by deriving a necessary condition set by the expansion of the universe. Specifically, we compare the stochastic acceleration timescale with the Hubble time, which defines the minimum requirement for particle energization on cosmological timescales. This approach provides an optimistic estimate of the acceleration efficiency and serves as a baseline against which additional physical effects, such as collisional energy losses, can be evaluated in subsequent sections.
Throughout this work, we focus on ions, specifically protons, as the dominant baryonic component of the plasma, which sets the relevant mass and gyrofrequency scales for the acceleration process.

\subsection{Acceleration Timescale}

We adopt a stochastic acceleration scenario in which particles gain energy through repeated scattering off moving magnetic irregularities. In general, the acceleration timescale depends on the velocity fluctuations at the scale with which particles interact, i.e., at scales comparable to the particle mean free path $\lambda(E,z)$.
The acceleration timescale can therefore be expressed schematically as
\begin{equation}
t_{\rm acc}(E,z) \sim \left(\frac{c}{v_\lambda(z)}\right)^2 \frac{\lambda(E,z)}{c},
\label{eq1}
\end{equation}
where $v_\lambda(z)$ denotes the characteristic turbulent velocity at the scale $\lambda$.
In the present work, we adopt an optimistic approximation in which particles are assumed to sample velocity fluctuations comparable to the outer-scale turbulent velocity, $v_\lambda \sim v_{\rm tur}$. Under this assumption, the acceleration timescale reduces to
\begin{equation}
t_{\rm acc}(E,z) \sim
\left(\frac{c}{v_{\rm tur}(z)}\right)^2
\frac{\lambda(E,z)}{c},
\label{eq2}
\end{equation}
which should be regarded as a lower-bound estimate of the true acceleration time. In realistic turbulence, where $v_\lambda < v_{\rm tur}$ for $\lambda$ below the outer scale, the actual acceleration time is expected to be longer, thereby further suppressing particle acceleration.

The mean free path is determined by the effective scattering frequency,
\begin{equation}
\lambda(E,z) \approx \frac{v(E)}{\nu_{\rm eff}(z)}.
\label{eq3}
\end{equation}
The effective scattering rate is defined as
\begin{equation}
\nu_{\rm eff}(z) \approx \left(|\Delta|-2\beta^{-1}\right)^{3/2}\omega_{\rm ci},
\label{eq4}
\end{equation}
where $\Delta$ is the pressure anisotropy factor, $\beta$ is the plasma beta parameter, and $\omega_{\rm ci} = eB/m_i c$ is the ion gyrofrequency associated with the evolving magnetic field.
This expression characterizes the pitch-angle scattering rate induced by pressure-anisotropy-driven microinstabilities, and is most directly applicable
to particles with velocities comparable to the thermal speed. 
For higher-energy particles, including particles that may have already been injected into the suprathermal regime by other mechanisms, the effective scattering rate is expected to depend on particle rigidity and on the spectrum of magnetic
fluctuations, leading to an energy-dependent diffusion coefficient. 
We therefore do not regard Eq.~\eqref{eq4} as a universal scattering law for suprathermal or relativistic particles. 
Instead, it is used here as an optimistic normalization of the instability-mediated scattering rate, chosen to estimate the maximum possible efficiency of the specific stochastic acceleration channel considered in this work. 
If the scattering efficiency decreases with rigidity, the acceleration time would be longer than estimated here, the critical magnetic field would be larger, and the production of suprathermal particles would be further suppressed. 
Thus, the results derived from Eq.~\eqref{eq4} should be interpreted as upper limits on the efficiency of this stochastic acceleration mechanism, rather than as generic predictions for all possible particle-injection channels.

Substituting Eq.~\eqref{eq3} into Eq.~\eqref{eq2}, the acceleration timescale becomes
\begin{equation}
t_{\rm acc}(z) \sim
\left(\frac{c}{v_{\rm tur}(z)}\right)^2
\frac{v(E)}{c\nu_{\rm eff}(z)}.
\label{eq5}
\end{equation}
This expression highlights that magnetic-field amplification, through its impact on $\omega_{\rm ci}$ and $\nu_{\rm eff}$, can significantly shorten the acceleration time.
To determine whether stochastic acceleration is viable before the formation of large-scale shocks, we compare the acceleration timescale with the Hubble time,
\begin{equation}
t_H(z) = H^{-1}(z),
\label{eq6}
\end{equation}
where
\begin{equation}
H(z) = H_0
\sqrt{\Omega_m(1+z)^3 + \Omega_\Lambda}.
\label{eq7}
\end{equation}
A necessary condition for efficient cosmological acceleration is therefore
\begin{equation}
t_{\rm acc}(E,z) < t_H(z).
\label{eq8}
\end{equation}
Combining the above expressions yields the dimensionless ratio
\begin{equation}
\frac{t_{\rm acc}}{t_H}
\approx
\left(\frac{c}{v_{\rm tur}}\right)^2
\left(\frac{v(E)}{c}\right)
\frac{H(z)}{\nu_{\rm eff}(z)}.
\label{eq9}
\end{equation}
When this ratio falls below unity, particles can be accelerated to relativistic energies within a cosmological expansion time. Conversely, if $t_{\rm acc} > t_H$, the expansion of the universe limits the achievable particle energies.

Eq.~\eqref{eq8} provides a necessary condition for stochastic acceleration on cosmological timescales. However, it is not sufficient, as additional energy loss processes can impose more stringent constraints, particularly at low energies.
In particular, the Coulomb energy loss rate can be comparable to or exceed the stochastic acceleration rate, $t_C^{-1} \gtrsim t_{\rm acc}^{-1}$, thereby suppressing efficient energization of sub-relativistic particles.
A more general requirement for particle acceleration is therefore
\begin{equation}
t_{\rm acc}(E,z) < \min\left[t_H(z), t_C(E,z)\right],
\end{equation}
where $t_C \equiv E/|\dot{E}_C|$ is the characteristic Coulomb energy loss timescale. This condition highlights that, even in the presence of efficient pitch-angle scattering, stochastic acceleration can be strongly suppressed at low energies due to collisional cooling. In this regime, particles undergo frequent scattering but gain energy only slowly, while Coulomb losses efficiently remove energy, preventing their injection into the nonthermal population.
In the present analysis, we first use the comparison with the Hubble time to identify the cosmological regime where acceleration may become viable. The impact of Coulomb losses is incorporated in Section~\ref{sec:s3}, where the acceleration condition is refined to include energy-dependent cooling effects that introduce an effective low-energy threshold for stochastic acceleration.

\subsection{Implications for Pre-Shock CR Production}

Eqs.~\eqref{eq1}-\eqref{eq9} suggest that instability-driven scattering during magnetogenesis may create a channel for CR production prior to structure formation shocks. As the magnetic field grows, the associated increase in gyrofrequency enhances the scattering rate, thereby reducing the acceleration timescale. This feedback implies that even moderately amplified magnetic fields could enable stochastic acceleration in the early universe.

In this framework, the key quantity controlling the onset of acceleration is the ratio $H(z)/\nu_{\rm eff}(z)$. If instability-mediated scattering becomes sufficiently strong, stochastic acceleration may operate well before the emergence of collisionless shocks, potentially establishing a seed population of CRs that later participate in DSA.
The results presented here therefore motivate a quantitative exploration of particle acceleration during cosmological magnetogenesis and its implications for the origin of the earliest nonthermal particle populations.
To evaluate the viability of stochastic acceleration  during pressure-anisotropy-driven magnetogenesis, we adopt a minimal  self-consistent closure of the background plasma parameters. 

We parametrize the turbulent velocity as a fixed fraction of the sound speed,
\begin{equation}
v_{\rm tur}(z) = \epsilon\, c_s(z),
\label{eq:vtur_param}
\end{equation}
where $\epsilon < 1$ is a dimensionless parameter and
\begin{equation}
c_s(z) = \sqrt{\frac{\gamma k_B T(z)}{\mu m_p}}
\end{equation}
is the adiabatic sound speed of the background plasma. We adopt $\mu \approx 0.59$ appropriate for a fully ionized primordial plasma.
Observations and simulations of diffuse astrophysical plasmas, such as the intracluster medium, suggest that gas motions are typically subsonic relative to the sound speed. For example, the {\it Hitomi} satellite measured a velocity dispersion of $\sim160,{\rm km,s^{-1}}$ in the Perseus cluster core, indicating subsonic turbulence \citep{Hitomi2016}, while numerical studies likewise find turbulent velocities that are a fraction of the sound speed \citep{Gaspari2013}. Although these results pertain to gas that has been processed by structure-formation shocks in the low-redshift Universe, they provide a useful reference for the characteristic magnitude of turbulent motions in weakly magnetized plasmas. Motivated by this, we adopt a representative range $\epsilon = 0.05$--$0.5$ in the present study.

We assume that the pressure anisotropy remains near a quasi-stable value
\begin{equation}
\Delta(z) \approx \Delta_0 = \text{const}.
\end{equation}
Near the marginal stability condition, the plasma beta satisfies
\begin{equation}
\beta(z) \sim \frac{2}{|\Delta_0|},
\label{eq:beta_closure}
\end{equation}
which provides a closure relation between $\beta$ and the anisotropy amplitude. Taking into account the system for particle acceleration, the anisotropy factor of the system $\Delta = \Delta_0 + \delta \Delta$ satisfies $|\Delta| > 2 \beta^{-1}$, where $\delta \Delta \ll \Delta_0$ represents a perturbation around the quasi-stable value $\Delta_0$, which is treated as a free parameter that drives the instability.
The supercritical excess above the marginal threshold $|\Delta_0| \sim 2\beta^{-1}$ is
\begin{equation}
\Delta_{\rm ex}(\delta,\beta)\equiv |\Delta_0+\delta \Delta|-|\Delta_0|
=|\Delta_0+\delta \Delta|-2\beta^{-1},
\label{eq:Delta_ex}
\end{equation}
which reduces to $\Delta_{\rm ex}=\delta \Delta$ for $\Delta_0>0$ and $\delta \Delta\ge 0$.
In high-$\beta$ astrophysical plasmas such as the intracluster medium (ICM), where $\beta \sim \mathcal{O}(10^2)$ \citep[e.g.,][]{Schekochihin2005, Schekochihin2006a, Schekochihin2006b, Ryu2008, Kunz2011, Kunz2022}, this corresponds to a typical anisotropy level $\Delta_0 \sim \mathcal{O}(10^{-2})$.

The magnetic field evolves according to the instability-driven growth equation derived in the scattering-dominated regime \citep{Falceta2015, Ha2025},
\begin{equation}
\frac{1}{B_0}\frac{dB}{dt}
\approx
\omega_{\rm ci,0}
\frac{\left(|\Delta|-2\beta^{-1}\right)^{3/2}}{1+|\Delta|}
\left(\frac{B}{B_0}\right)^2 \nonumber \
+
\omega_{\rm ci,0}
\frac{|\delta\Delta|^{3/2}}{1+|\Delta|}
e^{-\Gamma_d t-\frac{1}{2}\Gamma_d/\Gamma_c^{(\rm inst)}}
\left(\frac{B}{B_0}\right)^2 .
\end{equation}
The first term represents the baseline magnetic-field amplification driven by a sustained pressure anisotropy $\Delta$ in the scattering-dominated regime. When the anisotropy exceeds the instability threshold $|\Delta| > 2\beta^{-1}$, microinstabilities (e.g., mirror or firehose modes) generate magnetic fluctuations that enhance pitch-angle scattering, leading to a self-consistent growth of the magnetic field.
The second term describes the transient contribution associated with perturbations of the pressure anisotropy, $\delta\Delta$. This term accounts for additional magnetic-field amplification triggered by fluctuations in the anisotropy, but is exponentially suppressed over time due to damping of the anisotropy by pitch-angle scattering, characterized by the rate $\Gamma_d$. The factor involving $\Gamma_c^{(\rm inst)}$ represents the competition between anisotropy generation by large-scale plasma motions and its relaxation by instabilities.
Here, $B_0$ is the seed magnetic field and $\omega_{\rm ci,0}$ is the ion gyrofrequency evaluated at $B_0$. The quadratic dependence on $B$ reflects the nonlinear nature of the instability-driven amplification, implying that once the instability condition is satisfied, the magnetic field can grow rapidly.

In high-$\beta$ weakly magnetized plasmas,
pressure anisotropy is expected to be regulated near the instability
threshold by rapid pitch-angle scattering, so that the
strong-damping limit ($\Gamma_d \gg \Gamma_c^{(\rm inst)}$)
provides an appropriate closure.
For small perturbations of the anisotropy
($|\delta\Delta/\Delta| \ll 1$),
the perturbation-driven contribution becomes negligible.
While the detailed magnetic-field growth timescale can depend on
the relative magnitude of the damping and perturbation terms
\citep{Ha2025}, these effects primarily modify the growth rate
without significantly altering the overall magnetic-field evolution.
The magnetic-field evolution therefore reduces to
\begin{equation}
\frac{1}{B_0}\frac{dB}{dt}
\approx
\omega_{\rm ci,0}
\frac{\left(|\Delta|-2\beta^{-1}\right)^{3/2}}{1+|\Delta|}
\left(\frac{B}{B_0}\right)^2 .
\label{eq:B_evolution_closure}
\end{equation}
This nonlinear growth implies that once the instability condition
$|\Delta|>2\beta^{-1}$ is satisfied, the magnetic field amplifies
superlinearly with $B$, leading to rapid amplification during the
magnetogenesis phase.

\subsection{Analytic constraint on the critical magnetic field}
\label{subsec:Bcrit}

A necessary condition for efficient cosmological acceleration is that the acceleration time be shorter
than the Hubble time, $t_{\rm acc}<t_H=H^{-1}(z)$. Combining
Eqs.~ \eqref{eq9}, and \eqref{eq:Delta_ex} yields an analytic lower bound on the magnetic field:
\begin{equation}
B > B_{\rm crit}(z)
\approx
\frac{m_ic}{e}
\left(\frac{c}{v_{\rm tur}(z)}\right)^2
\left(\frac{v(E)}{c}\right)
\frac{H(z)}{\Delta_{\rm ex}^{3/2}}.
\label{eq:Bcrit_general}
\end{equation}
This condition defines a critical magnetic field required for stochastic acceleration to compete with cosmological expansion, and therefore represents a necessary (but not sufficient) criterion for particle acceleration.
Eq.~(\ref{eq:Bcrit_general}) makes explicit the strong parametric dependence
$B_{\rm crit}\propto H(z)\,\epsilon^{-2}c_s^{-2}(z)\Delta_{\rm ex}^{-3/2}$:
unless the turbulent motions are sufficiently strong and/or the anisotropy is strongly supercritical,
the magnetic field required for $t_{\rm acc}<t_H$ becomes unrealistically large, implying that
pressure-anisotropy-driven instabilities alone are unlikely to produce a significant pre-shock CR population under generic cosmological conditions.

Using Eq.~(\ref{eq:Bcrit_general}), we can derive an explicit redshift dependence of the
critical field required for $t_{\rm acc}<t_H$.
At $z\gtrsim 1$ (matter-dominated era) we approximate
\begin{equation}
H(z)\approx H_0\sqrt{\Omega_m}\,(1+z)^{3/2}.
\end{equation}
Adopting a parametrized thermal history,
\begin{equation}
T(z)=T_0\left(\frac{1+z}{1+z_0}\right)^{\alpha},
\end{equation}
the sound speed satisfies $c_s^2(z)\propto T(z)\propto (1+z)^{\alpha}$.
Eq.~(\ref{eq:Bcrit_general}) then yields
\begin{equation}
B_{\rm crit}(z)\propto \frac{H(z)}{c_s^2(z)}
\propto (1+z)^{\frac{3}{2}-\alpha},
\end{equation}
or equivalently
\begin{equation}
B_{\rm crit}(z)=B_{\rm crit}(z_0)
\left(\frac{1+z}{1+z_0}\right)^{\frac{3}{2}-\alpha},
\label{eq}
\end{equation}
where $B_{\rm crit}(z_0)$ collects the normalization factors, including $\epsilon^{-2}$ and $\Delta_{\rm ex}^{-3/2}$.
The parameter $\alpha$ characterizes the thermal history of the background plasma through $T(z)\propto(1+z)^\alpha$. In the absence of heating processes, adiabatic expansion of non-relativistic particles leads to $T\propto a^{-2}$, corresponding to $\alpha=2$. However, in realistic cosmological environments, additional heating mechanisms—such as photoheating, residual Compton heating, and structure formation—can partially offset adiabatic cooling, resulting in a shallower temperature evolution with $\alpha<2$.
For this reason, it is reasonable to consider a range of thermal histories with $1\lesssim\alpha\lesssim2$. In the illustrative case $\alpha=1$, corresponding to a moderately heated medium, the critical field scales as $B_{\rm crit}(z)\propto(1+z)^{1/2}$, implying a slowly increasing threshold toward higher redshift. By contrast, for purely adiabatic evolution ($\alpha=2$), one obtains $B_{\rm crit}(z)\propto(1+z)^{-1/2}$.

\subsection{Definition of the CR turn-on redshift}
\label{subsec:zon}

A useful way to quantify when pressure-anisotropy-driven scattering can begin to enable cosmological stochastic acceleration is to define a ``CR turn-on redshift" $z_{\rm on}$. We define $z_{\rm on}$ as the epoch at which the stochastic acceleration time becomes comparable to the Hubble time,
\begin{equation}
\left.\frac{t_{\rm acc}(z)}{t_H(z)}\right|_{z=z_{\rm on}} = 1,
\qquad t_H(z)=H^{-1}(z).
\label{eq:zon_def}
\end{equation}
This condition provides a necessary criterion for stochastic acceleration on cosmological timescales. For $z>z_{\rm on}$ the expansion time is too short ($t_{\rm acc}>t_H$) and stochastic acceleration is inefficient, while for $z<z_{\rm on}$ acceleration may in principle become viable.
However, this condition is not sufficient, as additional energy loss processes-such as Coulomb losses—can impose more stringent constraints, particularly at low energies. Therefore, the $z_{\rm on}$ defined here should be interpreted as an upper bound on the redshift below which stochastic acceleration may operate.

\begin{figure*}
\centering
\includegraphics[width=1\textwidth]{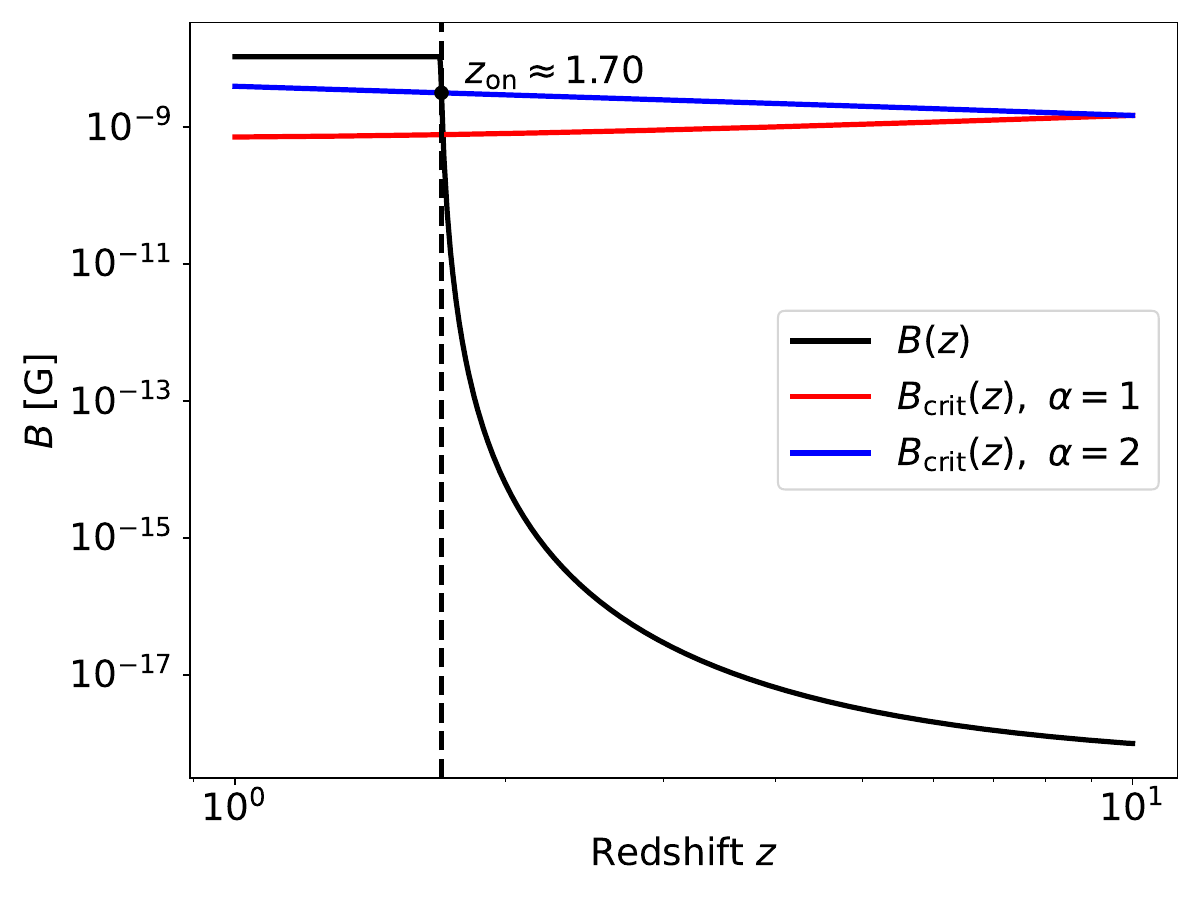}
\caption{
Redshift evolution of the magnetic field $B(z)$ (black) compared with the analytic critical field $B_{\rm crit}(z)$ for two different thermal histories: $\alpha=1$ (solid red) and $\alpha=2$ (dashed red), corresponding to a moderately heated medium and purely adiabatic evolution, respectively. Their intersection defines the CR turn-on redshift $z_{\rm on}$. The curves are shown for fiducial parameters $B_0=10^{-18}$~G, $\delta\Delta/\Delta_0=0.1$, $\Delta_0\sim2\beta^{-1}=10^{-2}$, and $\epsilon=0.2$. Here, the critical field is evaluated in the relativistic limit ($v\sim c$), providing a reference estimate for particle acceleration. The resulting $z_{\rm on}$ depends only weakly on the assumed thermal history.
}
\label{fig:f1}
\end{figure*}

\begin{figure*}
\centering
\includegraphics[width=1\textwidth]{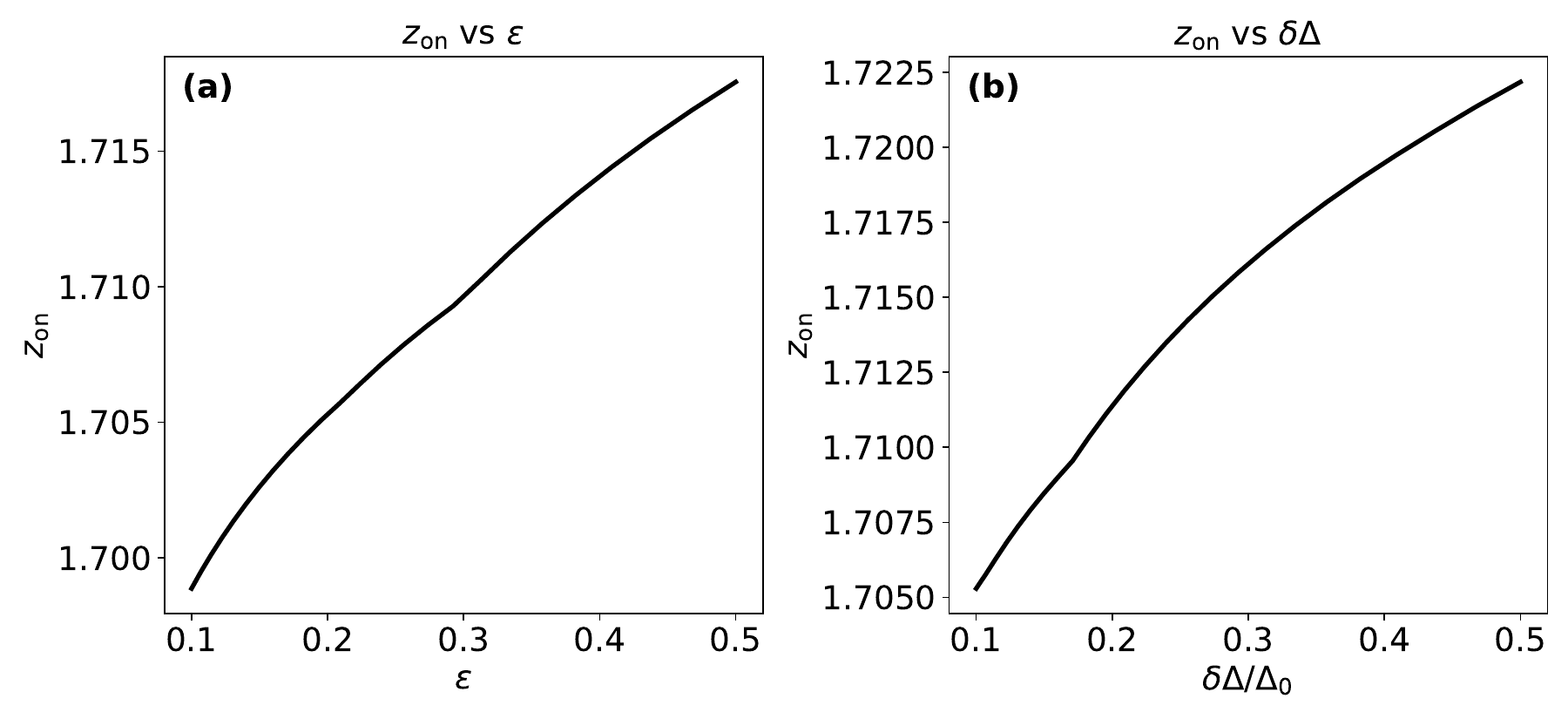}
\caption{
Dependence of the CR turn-on redshift $z_{\rm on}$ on
(a) the turbulence amplitude $\epsilon$ and
(b) the pressure-anisotropy perturbation $\delta\Delta$.
All other parameters are fixed to the fiducial values used in
Figure~\ref{fig:f1}.
Over the explored parameter range, $z_{\rm on}$ varies only weakly,
remaining close to $z_{\rm on}\sim1.7$.
}
\label{fig:f2}
\end{figure*}

The turn-on redshift $z_{\rm on}$ can be defined by the crossing between the evolving magnetic field and the critical field,
\begin{equation}
B(z_{\rm on}) = B_{\rm crit}(z_{\rm on}),
\label{eq:zon_crossing}
\end{equation}
which provides a convenient criterion for the onset of efficient stochastic acceleration.
In the matter-dominated regime, $H(z)\propto (1+z)^{3/2}$ and, for a parametrized thermal history $T(z)\propto (1+z)^{\alpha}$, the critical field scales as $B_{\rm crit}(z)\propto (1+z)^{3/2-\alpha}$. Assuming a simple power-law evolution of the magnetic field, $B(z)=B_0(1+z)^p$, the existence of a finite turn-on redshift depends on the relative scaling of these two quantities.
In particular, a crossing occurs only if the magnetic field grows sufficiently rapidly with redshift, $p>(3/2-\alpha)$; otherwise, $B(z)$ remains below $B_{\rm crit}(z)$ at all epochs, and stochastic acceleration remains inefficient.

To illustrate this behavior explicitly, we compute the magnetic-field evolution $B(z)$ using the unperturbed scattering-dominated solution from our previous work and compare it with the analytic critical field $B_{\rm crit}(z)$. Here, the critical field is evaluated in the relativistic limit ($v\sim c$), providing a baseline estimate for particle acceleration. Figure~\ref{fig:f1} shows the resulting redshift evolution for a representative set of parameters, including two different thermal histories characterized by $\alpha=1$ and $\alpha=2$.
At high redshift, the magnetic field remains far below the critical value in both cases, implying $t_{\rm acc}\gg t_H$ and therefore inefficient stochastic acceleration. As cosmic time progresses, nonlinear magnetic amplification driven by pressure-anisotropy instabilities leads to a rapid growth of $B(z)$. The CR turn-on redshift $z_{\rm on}$ is defined by the crossing $B(z)=B_{\rm crit}(z)$, marking the epoch at which stochastic acceleration becomes viable on cosmological timescales.
In all cases considered here, the crossing occurs at $z_{\rm on}\sim 1.7$, corresponding to a magnetic field strength of order $10^{-9}$~G. Notably, the location of this crossing depends only weakly on the assumed thermal history. This can be understood from the different redshift scalings of the two quantities: while $B_{\rm crit}(z)$ evolves only as a power law in $(1+z)$, the magnetic field undergoes rapid nonlinear amplification. As a result, the onset of stochastic acceleration is primarily governed by instability-driven magnetization, with the thermal history playing only a secondary role.
Importantly, this epoch coincides with the onset of nonlinear structure formation, suggesting that efficient cosmic-ray production is naturally associated with structure-formation shocks rather than with earlier microinstability-driven stochastic processes.

Figure~\ref{fig:f2} illustrates the dependence of the turn-on redshift on the turbulence amplitude $\epsilon$ and on the pressure anisotropy, parameterized by $\delta\Delta/\Delta_0$ and $\Delta_0 \sim 2\beta^{-1}$.
Despite the explicit appearance of both quantities in the analytic expression
for $B_{\rm crit}(z)$,
the resulting variation in $z_{\rm on}$ is remarkably small over the
considered parameter range.
In particular, even when $\epsilon$ and $\delta \Delta$ are varied by factors of a few,
the turn-on redshift remains confined to a narrow interval around
$z_{\rm on}\sim 1.7$.
This behavior reflects the fact that $B(z)$ exhibits a rapid nonlinear
growth phase, so that modest vertical shifts in $B_{\rm crit}(z)$
translate into only small horizontal shifts in the crossing redshift.
The onset of cosmologically viable stochastic acceleration is therefore
robust against moderate changes in the underlying microphysical parameters.
Importantly, the inferred $z_{\rm on}$ remains close to the epoch of nonlinear
structure formation, reinforcing the conclusion that efficient CR
production is naturally associated with structure-formation shocks rather
than with earlier microinstability-driven processes.

\subsection{Maximum attainable energy during cosmological magnetogenesis}
\label{subsec:Emax}

While the condition $t_{\rm acc}(z) < t_H(z)$ determines when stochastic
acceleration becomes cosmologically viable, it does not specify the maximum
particle energy that can be reached.
To quantify the physical significance of the turn-on epoch,
we estimate the maximum ion energy attainable before the expansion
of the Universe limits further acceleration.

To estimate the largest particle energies that could in principle be reached,  we adopt an optimistic strong-scattering estimate by assuming $\lambda \sim r_L$,  where $r_L$ is the Larmor radius. 
This corresponds to a Bohm-like limit,  $\nu_{\rm scatt}\sim \omega_{\rm ci}$, and should be interpreted as an illustrative upper bound rather than as a generic prediction of stochastic acceleration. 
In realistic turbulent media, the momentum diffusion coefficient and the acceleration timescale depend on the turbulence spectrum and on the nature of  the resonant scattering process. 
Different prescriptions, such as hard-sphere  diffusion, can therefore lead to different energy dependences of $t_{\rm acc}$.
In this limit, the spatial diffusion coefficient becomes
\begin{equation}
D(E) \sim \frac{1}{3} r_L(E)c,
\qquad
r_L(E) = \frac{pc}{eB} \approx \frac{E}{eB},
\label{eq:kappa_bohm_Emax}
\end{equation}
for relativistic ions with $E \approx pc$.
In this limit, the second-order Fermi acceleration time can be expressed as
\begin{equation}
t_{\rm acc,2}(E,z)
\approx
\left(\frac{c}{v_{\rm tur}(z)}\right)^2
\frac{r_L(E)}{c}
=
\left(\frac{c}{v_{\rm tur}(z)}\right)^2
\frac{E}{eB(z)c}.
\label{eq:tacc_E_dependence}
\end{equation}
This expression captures the essential energy dependence:
higher-energy particles require longer acceleration times.

\begin{figure*}
\centering
\includegraphics[width=1\textwidth]{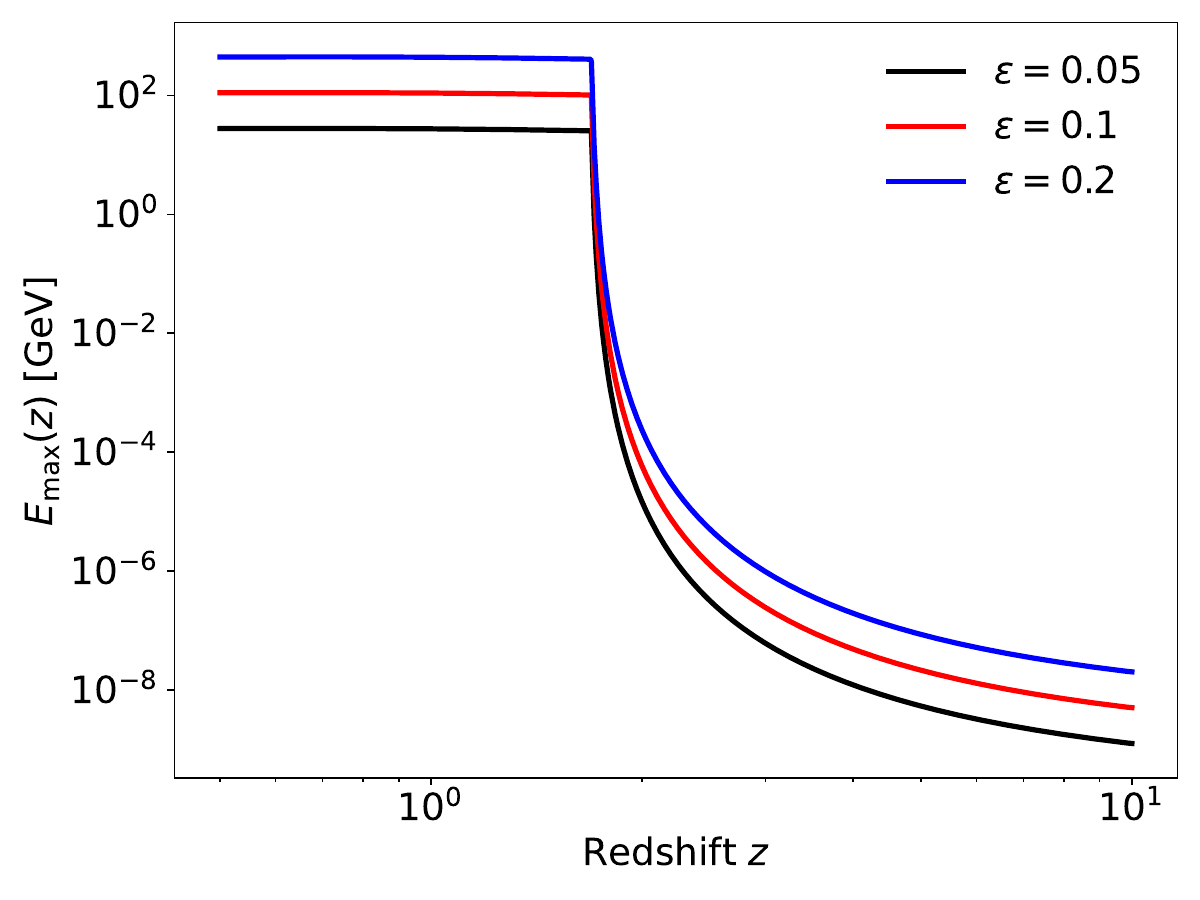}
\caption{
Redshift evolution of the maximum attainable ion energy
$E_{\max}(z)$ computed from Eq.~(\ref{eq:Emax_eps})
for turbulence amplitudes $\epsilon=0.05$ (black),
$0.1$ (red), and $0.2$ (blue).
Near the turn-on epoch $z_{\rm on}\sim1.7$,
$E_{\max}$ reaches $\sim10^2\,\mathrm{GeV}$.
}
\label{fig:f3}
\end{figure*}

The maximum attainable energy at a given redshift is obtained
by requiring that the acceleration time not exceed the Hubble time,
\begin{equation}
t_{\rm acc,2}(E_{\max},z) = t_H(z) = H^{-1}(z).
\label{eq:Emax_condition}
\end{equation}
Substituting the parametrization $v_{\rm tur}(z)=\epsilon c_s(z)$
into Eq.~(\ref{eq:tacc_E_dependence}), the maximum attainable energy becomes
\begin{equation}
E_{\max}(z)
\approx
e B(z) c
\epsilon^2
\left(\frac{c_s(z)}{c}\right)^2
H^{-1}(z).
\label{eq:Emax_eps}
\end{equation}
In the matter-dominated regime, $H(z)\propto (1+z)^{3/2}$.
Adopting the parametrized thermal history
$T(z)\propto (1+z)^{\alpha}$, so that
$c_s^2(z)\propto (1+z)^{\alpha}$,
we find
\begin{equation}
E_{\max}(z)
\propto
B(z)\,
(1+z)^{\alpha - 3/2}.
\label{eq:Emax_scaling}
\end{equation}
Thus, even if the magnetic field grows rapidly,
cosmological expansion suppresses the maximum achievable energy
at high redshift.

We emphasize that the scaling in Eq.~(\ref{eq:Emax_scaling}) follows from the Bohm-like assumption $\lambda\sim r_L$, for which the acceleration time increases  linearly with particle energy. 
It should therefore not be regarded as a universal energy scaling for stochastic acceleration.
In general, the momentum diffusion coefficient depends on the turbulence spectrum and scattering process \citep[e.g.,][]{Becker2006,Petrosian2012}.
If the momentum diffusion coefficient has a different energy dependence, the inferred $E_{\max}$ scaling will change. 
For example, a hard-sphere approximation would give an approximately energy-independent acceleration timescale, in which case the concept of $E_{\max}$ cannot be inferred from the simple linear scaling used here. 
Thus, the estimate below is intended only as an order-of-magnitude upper bound within the strong-scattering limit.

Figure~\ref{fig:f3} shows the redshift evolution of the maximum attainable ion energy for representative values of the turbulence parameter $\epsilon$.
Evaluated near the turn-on epoch $z_{\rm on}\sim 1.7$, where $B(z_{\rm on})\sim 10^{-9}\,\mathrm{G}$ in our fiducial model, Eq.~\eqref{eq:Emax_eps} implies that ion energies can, under optimistic assumptions corresponding to the strong-scattering limit, reach values up to $\sim 10^2\,\mathrm{GeV}$.
The strong dependence $E_{\max}\propto \epsilon^2$ implies that larger turbulence amplitudes lead to higher maximum energies, while the overall redshift dependence remains unchanged.
At higher redshift ($z\gtrsim z_{\rm on}$), the magnetic field remains weak and the Hubble time short, so that $E_{\max}$ is strongly suppressed and remains far below the relativistic regime.
Thus, even under optimistic assumptions, cosmological stochastic acceleration cannot generate a substantial high-energy particle population prior to the onset of nonlinear structure formation.
Consistent with the discussion above, Eq.~\eqref{eq:Emax_eps} should be interpreted as an order-of-magnitude upper-bound estimate within the Bohm-like strong-scattering limit. 
Additional effects, such as Coulomb and ionization losses, incomplete turbulence development, reduced scattering efficiency at high rigidity, or deviations from Bohm-like diffusion, would lower the attainable energy further. 
Thus, even if stochastic acceleration can in principle produce $\mathcal{O}(10$--$10^2)\,\mathrm{GeV}$ ions near $z_{\rm on}\sim 1.7$ in this limiting case, structure-formation shocks remain the more efficient and robust mechanism for accelerating CRs to higher energies.

\section{Refined Constraints Including Coulomb Losses in a Fully Ionized Medium}
\label{sec:s3}

In Section~\ref{sec:s2}, we derived the conditions for stochastic acceleration by considering adiabatic losses associated with cosmological expansion. 
This provides an optimistic estimate of the acceleration efficiency, corresponding to a lower bound on the acceleration time and a minimal requirement for particle energization.
In this section, we refine the analysis by including Coulomb losses, which arise from interactions with background electrons and become important at sub-relativistic energies. 
We focus on a fully ionized medium, appropriate for the post-reionization intergalactic medium. Since the turn-on redshift derived in this work lies at $z_{\rm on}\sim1.7$, well below the epoch of reionization ($z\sim6$), this assumption is well justified. 
In such an environment, Coulomb interactions dominate the collisional energy losses, allowing us to isolate the impact of collisional cooling on stochastic acceleration in a physically well-defined and tractable regime. 

\subsection{Energy loss processes in the pre-structure intergalactic medium}

In addition to adiabatic losses due to cosmological expansion, charged particles in the pre-structure intergalactic medium experience energy losses through Coulomb interactions with free electrons.
The characteristic Coulomb energy loss timescale is defined as
\begin{equation}
t_{\rm C}(E,z) \equiv \frac{E}{|\dot{E}_{\rm C}(E,z)|}.
\end{equation}
Coulomb losses are most effective at sub-relativistic energies and scale inversely with the ambient density, $t_{\rm C} \propto n^{-1}(z)$. Since the baryon density evolves as $n(z)\propto(1+z)^3$, the corresponding cooling times decrease rapidly toward higher redshift.
In ionized regions of the intergalactic medium, Coulomb interactions with free electrons dominate the collisional energy losses. Coulomb interactions of ions are generally dominated by scattering off thermal electrons \citep[e.g.,][]{Schlickeiser2002}, owing to the cumulative effect of frequent small-angle encounters with light particles. Accordingly, we approximate the Coulomb energy loss rate using the electron contribution.

Ionization losses can in principle contribute in partially neutral environments. However, in the present work we adopt a fully ionized medium as a fiducial model, appropriate for the post-reionization universe, and therefore neglect ionization losses. Including ionization processes would further shorten the cooling timescale at low energies, and thus strengthen the suppression of stochastic acceleration derived in this work.

For sub-relativistic and mildly relativistic ions in an ionized plasma, the Coulomb energy loss rate is dominated by interactions with background electrons and can be approximated as
\begin{equation}
-\dot{E}_{\rm C}(E,z)
\approx
\frac{4\pi e^4 n_e(z)\ln\Lambda}{m_e v(E)},
\label{eq:Coulomb_loss}
\end{equation}
where $n_e(z)$ is the electron number density, $\ln\Lambda$ is the Coulomb logarithm, and $v(E)$ is the particle velocity. This leads to a characteristic Coulomb loss timescale
\begin{equation}
t_{\rm C}(E,z)
\approx
\frac{Em_e v(E)}{4\pi e^4 n_e(z)\ln\Lambda}.
\end{equation}
This expression explicitly shows that Coulomb losses become increasingly efficient at low energies due to the inverse dependence on particle velocity.

The total effective loss rate is given by
\begin{equation}
t_{\rm loss}^{-1}(E,z)=t_H^{-1}(z)+t_C^{-1}(E,z),
\end{equation}
so that in the limits $t_C \ll t_H$ or $t_H \ll t_C$, the effective loss timescale reduces to the shorter of the two.
We note that the inclusion of ionization losses in partially neutral regions would further shorten the cooling timescale at low energies, thereby strengthening the constraints derived below.

\subsection{Refined critical magnetic field}

Including energy losses, the condition for stochastic acceleration becomes
\begin{equation}
t_{\rm acc}(E,z) < t_{\rm loss}(E,z),
\end{equation}
which defines an effective critical magnetic field $B_{\rm crit}^{\rm eff}(E,z)$ as the value at which the two timescales become comparable. Using the optimistic acceleration timescale from Section \ref{sec:s2}, we obtain
\begin{equation}
B_{\rm crit}^{\rm eff}(E,z) \approx
\frac{m_i c}{e}
\left(\frac{c}{v_{\rm tur}(z)}\right)^2
\left(\frac{v(E)}{c}\right)
\frac{1}{\Delta_{\rm ex}^{3/2} t_{\rm loss}(E,z)}.
\label{eq:Bcrit_eff}
\end{equation}
Since $t_{\rm loss}(E,z) \le t_H(z)$, energy losses increase the required magnetic field for efficient acceleration, $B_{\rm crit}^{\rm eff}(E,z) \ge B_{\rm crit}(z)$. In the Coulomb-dominated regime ($t_{\rm loss} \approx t_C$), this introduces an explicit energy dependence, implying that low-energy particles are more strongly suppressed by collisional cooling.

\begin{figure*}
\centering
\includegraphics[width=1\textwidth]{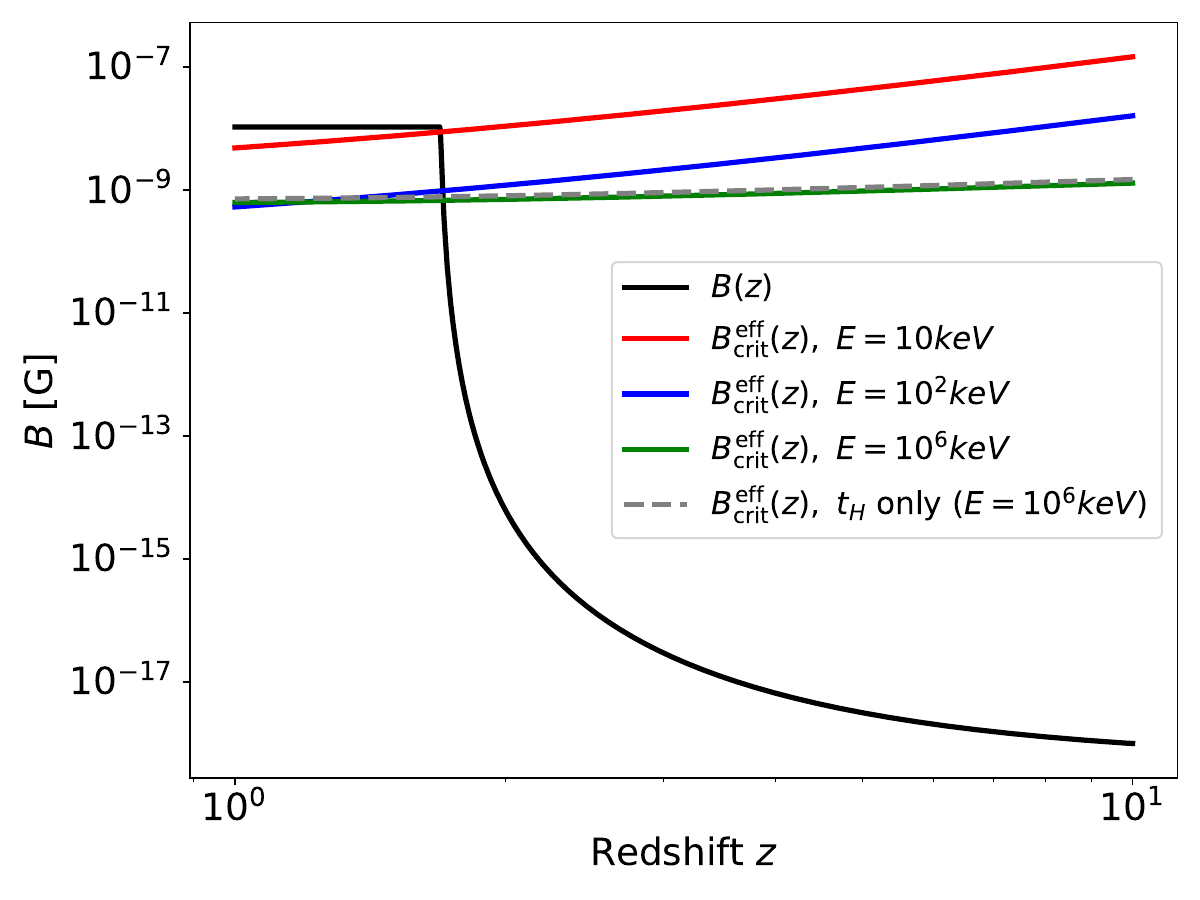}
\caption{
Redshift evolution of the magnetic field $B(z)$ (black) compared with the refined critical field $B_{\rm crit}^{\rm eff}(E,z)$ for different particle energies including Coulomb losses. The gray curve shows the Hubble-limited result. Coulomb losses strongly suppress acceleration at low energies, while at high energies $B_{\rm crit}^{\rm eff}$ converges to the Hubble-limited case, indicating an effective low-energy threshold for stochastic acceleration.
}
\label{fig:f4}
\end{figure*}

Figure~\ref{fig:f4} shows the redshift evolution of the refined critical magnetic field $B_{\rm crit}^{\rm eff}(E,z)$ for a range of particle energies, together with the magnetic-field evolution $B(z)$ and the Hubble-limited critical field.
At low energies, Coulomb losses significantly reduce the effective loss timescale ($t_{\rm loss}\approx t_{\rm C}$), leading to a substantial increase in $B_{\rm crit}^{\rm eff}$. As a result, the required magnetic field for efficient stochastic acceleration becomes much larger than in the adiabatic-only case, and the crossing with $B(z)$ is either shifted to much lower redshift or entirely suppressed. This demonstrates that collisional cooling strongly inhibits particle acceleration near the injection scale.
In contrast, at sufficiently high energies where $t_{\rm C}\gg t_H$, the effective loss timescale is governed by cosmological expansion ($t_{\rm loss}\approx t_H$), and $B_{\rm crit}^{\rm eff}$ converges to the Hubble-limited critical field derived in Section~\ref{sec:s2}. In this regime, stochastic acceleration becomes largely insensitive to Coulomb losses, and the corresponding turn-on redshift approaches the relativistic baseline shown in Figure~\ref{fig:f1}.
These results clearly demonstrate that Coulomb losses introduce a strong energy dependence in the conditions for stochastic acceleration, primarily affecting low-energy particles while leaving high-energy particles largely unaffected.

The comparison between $B(z)$ and $B_{\rm crit}^{\rm eff}(E,z)$ further implies the existence of an effective low-energy threshold for stochastic acceleration. Below this threshold, Coulomb losses dominate over acceleration, preventing particles from gaining energy efficiently and suppressing their injection into the acceleration process.
In the fiducial model considered here, this threshold is found to lie at approximately the ${\mathcal O}(10)$ keV level. Particles with energies below this scale experience rapid collisional cooling, such that $t_{\rm C} \ll t_{\rm acc}$ over the relevant redshift range. As a result, stochastic acceleration is effectively quenched. Conversely, for particles with energies above this threshold, the Coulomb loss timescale increases rapidly, allowing stochastic acceleration to operate and eventually approach the Hubble-limited regime.
We therefore conclude that Coulomb losses impose an effective low-energy threshold for stochastic acceleration, transforming the acceleration condition from a purely cosmological constraint into an energy-dependent threshold problem. At sufficiently high energies, however, where $t_{\rm C}\gg t_H$, the acceleration process becomes limited by cosmological expansion, and the maximum attainable energy remains essentially unchanged from the adiabatic-only case.

\section{Pre-Structure Stochastic Acceleration in the Early Universe}
\label{sec:s4}

To quantify the magnitude of any seed nonthermal population produced by the instability-assisted stochastic process discussed above, we introduce a transport model for the isotropic ion distribution function. 
The goal is not to construct a fully self-consistent cosmological CR population synthesis, but rather to translate the acceleration-time constraints into an explicit estimate of the resulting momentum-space spectrum and energy content.

\subsection{Isotropic Fokker-Planck equation in an expanding background}

Let $f(p,z)$ be the isotropic phase-space distribution of ions,
normalized such that the differential number density is
$dn = 4\pi p^{2} f(p,z)\,dp$.
Neglecting spatial gradients and assuming isotropic
pitch-angle scattering, the momentum-space Fokker-Planck equation reads
\begin{equation}
\frac{\partial f}{\partial t}
=
\frac{1}{p^{2}}
\frac{\partial}{\partial p}
\left[
p^{2} D_{pp}(p,z)\,\frac{\partial f}{\partial p}
\right] 
-
\frac{1}{p^{2}}
\frac{\partial}{\partial p}
\left[
p^{2}\,\dot p_{\rm loss}(p,z)\, f
\right]
+Q(p,z)
-\frac{f}{t_{\rm esc}(p,z)}, \,
\label{eq:fp_time}
\end{equation}
where $D_{pp}$ is the momentum diffusion coefficient associated with
second-order Fermi acceleration, $\dot p_{\rm loss}$ represents systematic
momentum losses, $Q$ is a source term (e.g., injection from the thermal pool),
and $t_{\rm esc}$ is an effective escape time from the acceleration region.
Cosmological expansion is incorporated by transforming from $t$ to redshift $z$
using
\begin{equation}
\frac{dt}{dz} = -\frac{1}{(1+z)H(z)} \, ,
\label{eq:dtdz}
\end{equation}
with $H(z)=H_0\sqrt{\Omega_{\rm m}(1+z)^3+\Omega_\Lambda}$.
Eq.~\eqref{eq:fp_time} then becomes
\begin{equation}
-(1+z)H(z)\,\frac{\partial f}{\partial z}
=
\frac{1}{p^{2}}
\frac{\partial}{\partial p}
\left[
p^{2} D_{pp}(p,z)\,\frac{\partial f}{\partial p}
\right]
-
\frac{1}{p^{2}}
\frac{\partial}{\partial p}
\left[
p^{2}\,\dot p_{\rm loss}(p,z)\, f
\right]
+Q(p,z)\, .
\label{eq:fp_z}
\end{equation}

In a second-order Fermi process associated with the pressure-anisotropy-driven instabilities, which generate small-scale magnetic fluctuations that scatter particles, the characteristic acceleration time can be expressed in terms of $D_{pp}$ as
\begin{equation}
D_{pp}(p,z)=\frac{p^{2}}{t_{\rm acc}(z)}.
\label{eq:Dpp_closure}
\end{equation}
This corresponds to a momentum diffusion coefficient proportional to $p^{2}$,
with a redshift-dependent normalization determined by the same stochastic
acceleration timescale used in the analytic estimates of Section~\ref{sec:s2}.
Thus, Eq.~\eqref{eq:Dpp_closure} provides a direct link between the timescale
criterion and the Fokker--Planck calculation.

The purpose of the Fokker--Planck treatment is not to assume a separate pre-acceleration mechanism or an externally imposed injection momentum.
Rather, it tests whether stochastic momentum diffusion, acting on an initially thermal or weakly suprathermal distribution, can overcome adiabatic and Coulomb losses and produce a nonthermal tail. 
The effective low-energy threshold discussed below therefore emerges from the competition between stochastic diffusion and Coulomb cooling, rather than being imposed as an external boundary condition.

In the present model, we assume that the escape timescale is much longer than the characteristic acceleration and loss timescales, $t_{\rm esc} \gg t_{\rm acc},~t_{\rm loss}$. The characteristic loss timescale is defined as $t_{\rm loss} \equiv p/\dot{p}_{\rm loss}$.
Cosmological expansion inevitably induces adiabatic momentum losses,
\begin{equation}
\dot p_{\rm ad}(p,z) \approx -H(z)p.
\label{eq:pad}
\end{equation}
In addition, particles experience Coulomb losses due to interactions with background electrons, which are particularly important at sub-relativistic energies. The total momentum loss rate is therefore given by
\begin{equation}
\dot p_{\rm loss}(p,z) = \dot p_{\rm ad}(p,z) + \dot p_{\rm C}(p,z),
\end{equation}
where the Coulomb momentum loss rate can be approximated as
\begin{equation}
\dot p_{\rm C}(p,z)
\approx
-\frac{4\pi e^4 n_e(z)\ln\Lambda}{m_e v^2(p)}.
\label{eq:dpC}
\end{equation}
Here $n_e(z)$ is the electron number density, $\ln\Lambda$ is the Coulomb logarithm, and $v(p)$ is the particle velocity. This expression shows that Coulomb losses increase rapidly toward low velocities, leading to efficient thermalization of particles near the injection scale.
Including both contributions, Eq.~\eqref{eq:fp_z} becomes
\begin{equation}
-(1+z)H(z)\frac{\partial f}{\partial z}
\approx
\frac{1}{p^{2}}
\frac{\partial}{\partial p}
\left[
p^{2}\frac{p^{2}}{t_{\rm acc}(z)}\frac{\partial f}{\partial p}
\right]
-
\frac{1}{p^{2}}
\frac{\partial}{\partial p}
\left[
p^{2}\left(\dot p_{\rm ad}(p,z) + \dot p_{\rm C}(p,z)\right) f
\right]
+Q(p,z).
\label{eq:fp_final_coulomb}
\end{equation}
The Coulomb term introduces an additional sink of particle momentum at low energies, which can suppress the formation of a suprathermal tail by efficiently thermalizing particles near the injection scale.

We note that adiabatic momentum losses due to cosmological expansion are applicable to particles residing in the unbound intergalactic medium considered in this work. In contrast, particles confined within gravitationally bound systems, such as galaxies or galaxy clusters, do not experience cosmological adiabatic cooling. Since our analysis focuses on the pre-structure phase prior to nonlinear collapse, the assumption $\dot p_{\rm ad} \approx -H(z)p$ remains valid throughout the redshift interval of interest.

\subsection{Numerical implementation and early-universe heating}

To assess the magnitude of stochastic heating prior to the magnetic-field turn-on epoch, we solve Eq.~\eqref{eq:fp_final_coulomb} numerically over the redshift interval $z=10 \rightarrow z_{\rm on}$, where the magnetic field remains well below its rapid-amplification regime. In this early phase, the goal is to determine whether the instability-assisted second-order Fermi process can produce a significant seed nonthermal component over a Hubble time.

The initial condition at $z=10$ is taken to be a non-relativistic Maxwellian distribution,
\begin{equation}
f_{\rm MW}(p,z=10) =
\frac{n_i}{(2\pi m_i k_B T_0)^{3/2}}
\exp\left(-\frac{p^{2}}{2m_i k_B T_0}\right),
\end{equation}
where $n_i$ denotes the ion number density. We emphasize that this setup is not intended to represent the volume-averaged thermal state of the universe at $z=10$, but rather a simplified diffuse ionized medium used as a fiducial environment for assessing the competition between stochastic acceleration and cooling processes. Accordingly, $T_0$ should be interpreted as an effective model parameter. In the fiducial case, we adopt $k_B T_0 = 0.86~{\rm eV}$, corresponding to a moderately heated plasma.
No explicit source term is included ($Q=0$), and escape is neglected, so that the evolution is governed solely by stochastic momentum diffusion and energy losses. The magnetic field $B(z)$ is mapped to redshift using the cosmological relation $dt/dz=-[(1+z)H(z)]^{-1}$. In the early regime ($z\gtrsim z_{\rm on}$), $B(z)$ evolves smoothly and remains far below the explosive growth associated with $z<z_{\rm on}$, ensuring that $t_{\rm acc}(z)$ varies gradually.

Eq.~\eqref{eq:fp_final_coulomb} is solved on a logarithmic momentum grid
using a fully implicit scheme in redshift.
The equation is written in conservative flux form in momentum space,
\begin{equation}
\mathcal{R}(f)
=
\frac{1}{p^2}
\frac{\partial}{\partial p}
\left[
p^2 D_{pp} \frac{\partial f}{\partial p}
-
p^2 (\dot p_{\rm ad} + \dot p_{\rm C}) f
\right],
\end{equation}
and discretized such that particle number conservation is maintained
up to numerical accuracy.
The redshift step $\Delta z$ is mapped to an effective time step
$\Delta t = -\Delta z /[(1+z)H(z)]$,
ensuring consistency with the cosmological evolution.

\begin{figure*}
\centering
\includegraphics[width=1\textwidth]{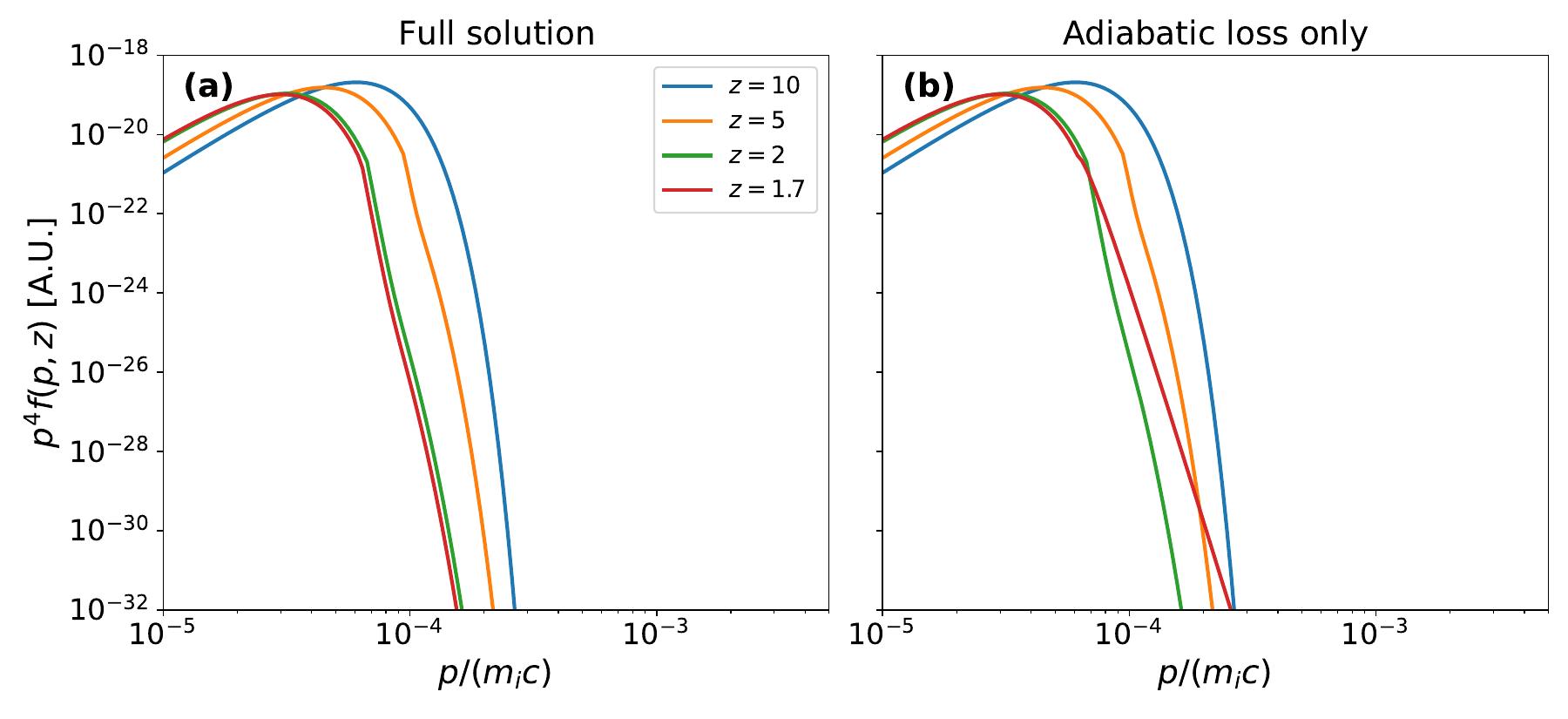}
\caption{
Ion spectra $p^{4}f(p,z)$ from the Fokker--Planck model at $z=10,5,2,$ and $1.7$. Panel (a): full solution including Coulomb and adiabatic losses; panel (b): adiabatic-loss-only case. The initial distribution at $z=10$ is a Maxwellian with $k_B T_0=0.86~{\rm eV}$. The turbulent velocity is parameterized as $v_{\rm tur}=\epsilon c_s$ with $\epsilon=0.2$. Coulomb losses suppress the formation of a suprathermal tail by efficiently cooling low-energy particles.
}
\label{fig:f5}
\end{figure*}

Figure~\ref{fig:f5} shows the evolution of the particle spectra $p^4 f(p,z)$ at
$z=10,5,2,$ and $1.7$, comparing the full solution including Coulomb losses (left panel) with the adiabatic-loss-only case (right panel).
In both cases, the thermal peak shifts toward lower momentum as redshift
decreases, reflecting adiabatic cooling for which
$p \propto a^{-1} \propto (1+z)$.
The corresponding decrease in peak amplitude is likewise consistent
with the dilution of the thermal energy density in an expanding background.
In the absence of Coulomb losses (right panel), stochastic momentum diffusion produces a modest broadening of the high-momentum tail relative to pure adiabatic evolution.
For stronger turbulence, the suprathermal component becomes slightly more pronounced as the system approaches the magnetic-field turn-on epoch, indicating that stochastic acceleration begins to compete with adiabatic cooling.
In contrast, when Coulomb losses are included (left panel), the low-energy population is efficiently thermalized, and the formation of a suprathermal tail is strongly suppressed over the entire redshift range considered.
Even for relatively strong turbulence, the distribution remains close to a Maxwellian, with only minimal deviation at high momenta.
This behavior reflects the fact that Coulomb cooling dominates over stochastic acceleration near the injection scale, preventing particles from being efficiently accelerated out of the thermal pool.
The effective low-energy threshold inferred from this behavior is not imposed externally, but arises dynamically from the competition between Coulomb cooling and stochastic momentum diffusion.
As the system approaches the magnetic-field turn-on epoch
($z \sim z_{\rm on} \sim 1.7$), the acceleration timescale decreases due to magnetic-field amplification.
However, the results indicate that Coulomb losses continue to limit the development of a significant nonthermal component in the pre-shock phase.
We therefore conclude that stochastic acceleration alone is unable to generate a substantial suprathermal seed population prior to the onset of structure-formation shocks.

We do not extend the Fokker--Planck integration below $z_{\rm on}$, since in this regime large-scale structure formation shocks are expected to dominate particle acceleration.
The present model is therefore designed to quantify the pre-shock stochastic heating phase and to assess the extent to which a seed nonthermal population can be established before shock-driven acceleration takes over.

This conclusion applies only to the stochastic acceleration channel modeled here.
It does not exclude the possibility that a separate, faster injection mechanism, such as magnetic reconnection within a turbulent cascade, could generate suprathermal particles before the onset of structure-formation shocks.
Indeed, recent numerical studies have shown that reconnecting current sheets in turbulent plasmas can provide rapid particle injection \citep[e.g.,][]{Comisso2018,Comisso2019}, followed by further stochastic acceleration by turbulent fluctuations.
Including such a mechanism would require an additional source term or injection model in the transport equation, which is beyond the scope of the present work.

\subsection{Limited impact of a pre-existing suprathermal population}

While the preceding analysis assumes a purely thermal initial distribution, a weak suprathermal particle population may arise from a variety of processes operating in weakly magnetized plasmas, such as plasma instabilities, turbulence, or residual nonthermal components inherited from earlier evolutionary stages. Even a small nonthermal tail could in principle act as a seed population for subsequent acceleration.

To assess the sensitivity of the stochastic heating process to such initial conditions, we consider an alternative initial distribution consisting of a Maxwellian core with a small suprathermal power-law tail,
\begin{equation}
f(p,z=10) = f_{\rm MW}(p,z=10) + Ap^{-s}\exp(-p/p_{\rm cut}) .
\end{equation}
The normalization of the tail is fixed by matching it to the Maxwellian distribution at a reference momentum $p_0$, such that
\begin{equation}
A p_0^{-s} = \eta_{\rm tail} f_{\rm MW}(p_0,z=10),
\end{equation}
where $\eta_{\rm tail}\ll 1$ controls the relative amplitude of the suprathermal component. This condition ensures a smooth connection between the thermal and nonthermal populations at $p=p_0$.
We then evolve this distribution using Eq.~\eqref{eq:fp_final_coulomb}. The slope of the pre-existing tail is taken to be $s=4.2$, representative of typical nonthermal spectra in collisionless plasmas, and consistent with expectations from both stochastic acceleration and shock-related processes. An exponential cutoff $p_{\rm cut}$ limits the high-momentum extent of the seed population.

\begin{figure*}
\centering
\includegraphics[width=1\textwidth]{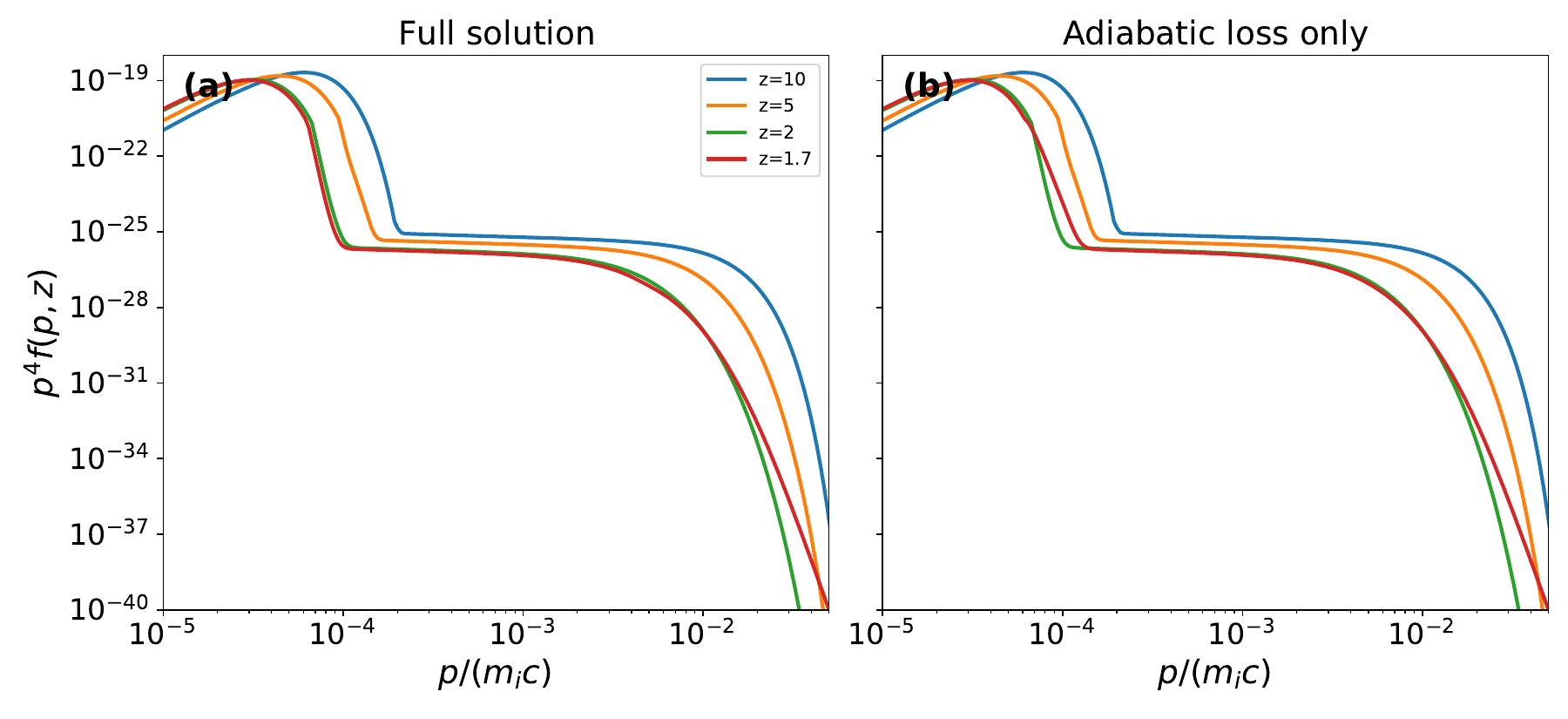}
\caption{
Same as Figure~\ref{fig:f5}, but including a weak pre-existing suprathermal power-law tail ($s=4.2$) in the initial distribution.
}
\label{fig:f6}
\end{figure*}

The resulting spectra are shown in Figure~\ref{fig:f6}.
We find that the overall evolution from $z=10$ to $z=2$ is largely
insensitive to the presence of a weak pre-existing suprathermal
population.
In both panels, the ion distribution shifts toward lower momenta as the universe expands, reflecting adiabatic momentum losses
$\dot p_{\rm ad}=-H(z)p$, for which $p\propto a^{-1}\propto(1+z)$.
However, when Coulomb losses are included, a clear departure from purely adiabatic evolution appears at low momenta.
In particular, particles with $p/(m_ic)\lesssim 5\times10^{-3}$,
corresponding to kinetic energies of order $\sim 10~{\rm keV}$,
are efficiently thermalized by Coulomb interactions.
In this regime, the cooling timescale is shorter than the acceleration
timescale, and stochastic diffusion is effectively suppressed.
As a result, the pre-existing power-law tail does not survive at low energies,
and the distribution remains close to a Maxwellian core throughout the evolution.
At higher momenta ($p/(m_ic)\gtrsim 5\times10^{-3}$),
the Coulomb loss timescale increases rapidly, allowing stochastic
acceleration to operate.
Nevertheless, even in this regime, the resulting modification of the
spectrum remains modest over the interval $z=10\rightarrow2$,
indicating that cosmological cooling continues to dominate the overall
evolution.

As the system approaches the magnetic-field turn-on epoch
($z\sim z_{\rm on}\sim1.7$), the reduction of the stochastic acceleration
timescale allows the diffusion term to partially compete with losses.
However, the results indicate that Coulomb cooling still limits the
efficient re-acceleration of the seed population.
The presence of a pre-existing tail therefore does not significantly
enhance the formation of a nonthermal component in the pre-shock phase.
Instead, the early evolution is controlled primarily by the competition
between Coulomb thermalization and cosmological expansion, with stochastic
acceleration becoming effective only above the effective threshold
set by Coulomb losses.

\section{Comparison with shock acceleration in the Bohm-diffusion limit}
\label{sec:s5}

Even if stochastic acceleration becomes formally viable after the turn-on epoch $z_{\rm on}$, it is important to assess whether it can compete with acceleration in structure-formation shocks.
To this end, we compare the characteristic acceleration timescale of our stochastic acceleration mechanism with the standard DSA timescale assuming Bohm-like diffusion.

For a non-relativistic shock of speed $u_s$ (in the upstream frame), the DSA acceleration time can be written
as \citep[e.g.,][]{Drury1983}
\begin{equation}
t_{\rm acc,sh}(E)
\approx
\frac{3}{u_1-u_2}\left(\frac{D_1(E)}{u_1}+\frac{D_2(E)}{u_2}\right),
\label{eq:tacc_dsa_general}
\end{equation}
where $u_{1,2}$ and $D_{1,2}$ are the upstream/downstream flow speeds and spatial diffusion
coefficients, respectively.
For a strong shock with compression ratio $r=u_1/u_2\approx 4$ and comparable diffusion coefficients on both sides
($D_1\sim D_2 \sim D$), Eq.~(\ref{eq:tacc_dsa_general}) reduces to the commonly used estimate
\begin{equation}
t_{\rm acc,sh}(E)\sim \xi\,\frac{D(E)}{u_s^2},
\qquad
\xi\sim \mathcal{O}(10),
\label{eq:tacc_dsa_kappa}
\end{equation}
where $\xi$ absorbs order-unity factors depending on the shock structure.
In the Bohm-like limit, the diffusion coefficient is
\begin{equation}
D_{\rm B}(E)\equiv \frac{1}{3}r_L(E)c,
\qquad
r_L(E)=\frac{pc}{eB}\approx \frac{E}{eB},
\label{eq:kappa_bohm}
\end{equation}
for relativistic particles with $E\approx pc$.
Combining Eqs.~(\ref{eq:tacc_dsa_kappa})--(\ref{eq:kappa_bohm}) yields
\begin{equation}
t_{\rm acc,sh}(E)\sim \frac{\xi}{3}\,\frac{r_L(E)c}{u_s^2}
\sim
\frac{\xi}{3}\,\frac{E}{eB}\,\frac{c}{u_s^2}.
\label{eq:tacc_sh_bohm}
\end{equation}

To compare the characteristic acceleration timescales of DSA and stochastic acceleration, we form the ratio
\begin{equation}
\frac{t_{\rm acc,2}}{t_{\rm acc,sh}}
\sim
\frac{3}{\xi}\,
\left(\frac{c}{v_{\rm tur}}\right)^2
\left(\frac{u_s}{c}\right)^2
\left(\frac{c}{r_L\,\nu_{\rm eff,B}}\right),
\label{eq:ratio_2nd_to_shock}
\end{equation}
where $\xi\sim \mathcal{O}(10)$ parameterizes order-unity factors in the DSA timescale.
The final factor, $c/(r_L\nu_{\rm eff,B})$, measures the efficiency of pitch-angle scattering
over a gyro-orbit; in the Bohm-like limit, $\nu_{\rm eff,B}\sim c/r_L$, this factor is of order unity.

For structure-formation environments, the relevant characteristic velocities are $v_{\rm tur}\sim \epsilon c_s$ with $c_s\sim 10^2-10^3~{\rm km\,s^{-1}}$ and $\epsilon\sim 0.1$, implying $v_{\rm tur}/c\sim10^{-4}-10^{-3}$.
Large-scale structure shocks typically have $u_s\sim10^3~{\rm km\,s^{-1}}$, corresponding to $u_s/c\sim3\times10^{-3}$.
Adopting Bohm-like scattering ($c/(r_L\nu_{\rm eff,B})\sim1$), these scalings imply
\begin{equation}
\frac{t_{\rm acc,2}}{t_{\rm acc,sh}}
\sim 10-10^{3},
\end{equation}
up to order-unity factors.
Thus, even under optimistic assumptions for the turbulence-driven stochastic mechanism, second-order acceleration is typically one to three orders of magnitude slower than shock acceleration in the same environment.

Consequently, once nonlinear structure formation generates shocks with
$u_s\sim 10^3~{\rm km\,s^{-1}}$ and simultaneously amplifies the magnetic field,
DSA operates on substantially shorter timescales.
Even if the instability-assisted stochastic channel formally turns on at
$z\lesssim z_{\rm on}$, shock acceleration is expected to dominate the production of CRs in that epoch.
This supports our central conclusion: a significant pre-shock CR population
is unlikely, and efficient CR production is naturally tied to
the onset of structure-formation shocks.

\section{Summary and Discussion}
\label{sec:s6}

We have investigated whether stochastic (second-order Fermi) acceleration associated with pressure-anisotropy-driven magnetogenesis can generate a dynamically significant population of cosmic rays (CRs) prior to the onset of nonlinear structure formation. By combining analytic acceleration-time constraints with a Fokker--Planck description of particle transport, we quantified both the conditions under which stochastic acceleration becomes cosmologically viable and the resulting ion momentum distribution in the pre-structure intergalactic medium.

An analytic comparison between the acceleration timescale and the Hubble time yields a critical magnetic field $B_{\rm crit}(z)$ and defines a characteristic turn-on redshift $z_{\rm on}\sim 1.7$, at which stochastic acceleration can in principle begin to compete with cosmological expansion. We find that this turn-on epoch is largely insensitive to uncertainties in the thermal history and microphysical parameters, reflecting the rapid nonlinear amplification of the magnetic field. However, this condition provides only an optimistic upper bound on the efficiency of particle acceleration.

When additional energy loss processes are included, the situation changes qualitatively. In particular, Coulomb interactions in a fully ionized intergalactic medium introduce a strong energy dependence in the effective loss timescale, leading to a refined critical magnetic field $B_{\rm crit}^{\rm eff}(E,z)$. This analysis reveals the existence of an effective low-energy threshold for stochastic acceleration at energies of order ${\cal O}(10)$ keV, corresponding to $p/(m_ic)\sim 5\times10^{-3}$. Below this scale, Coulomb losses dominate over acceleration ($t_{\rm C}\ll t_{\rm acc}$), efficiently thermalizing particles and suppressing their injection into the acceleration process. At higher energies, where $t_{\rm C}\gg t_H$, the acceleration condition approaches the Hubble-limited regime, and the maximum attainable energy remains essentially unchanged from the adiabatic-only estimate.

To quantify the resulting particle population, we solved a Fokker--Planck equation for the isotropic ion distribution over the redshift interval $z=10\rightarrow z_{\rm on}$. The numerical results show that, even under optimistic assumptions corresponding to the strong-scattering limit, the ion distribution remains close to a cooling Maxwellian throughout this epoch. Adiabatic expansion dominates the evolution, and Coulomb losses further suppress the formation of a suprathermal tail by efficiently removing particles near the injection scale. Even when a pre-existing suprathermal component is included in the initial condition, it is rapidly thermalized at low energies and does not lead to a significant enhancement of the nonthermal population.

These results imply that stochastic acceleration during pressure-anisotropy-driven magnetogenesis is unable to generate a dynamically important CR population in the pre-structure universe. In particular, Coulomb losses transform the acceleration problem from a purely cosmological constraint into an energy-dependent injection problem, strongly limiting the fraction of particles that can participate in stochastic energization. As a consequence, the resulting suprathermal population remains energetically subdominant and is unlikely to have a measurable impact on the thermal history of the intergalactic medium or on observable high-energy backgrounds.

Our findings therefore support a picture in which efficient CR production is intrinsically linked to the onset of nonlinear structure formation. While instability-driven scattering can enhance particle transport and may provide a modest level of pre-acceleration, the emergence of a substantial nonthermal population requires the presence of large-scale shocks and sufficiently amplified magnetic fields. The low-level suprathermal population obtained here may serve as a seed for subsequent diffusive shock acceleration (DSA), but its contribution to the overall CR energy budget is expected to be limited. We stress, however, that this conclusion does not rule out suprathermal seed formation by faster injection mechanisms. In particular, magnetic reconnection within turbulent current sheets may inject particles from the thermal pool into the suprathermal regime on timescales shorter than the stochastic acceleration timescale considered in this work. 
Incorporating such reconnection-driven injection would require an additional source term or injection prescription in the transport equation, and is left for future work together with a detailed kinetic study of particle injection and re acceleration at structure-formation shocks.

\printcredits

\bibliographystyle{cas-model2-names}

\bibliography{ref_stochastic_R1}



\end{document}